\documentclass[3p]{elsarticle}

\makeatletter
\def\ps@pprintTitle{%
  \let\@oddhead\@empty
  \let\@evenhead\@empty
  \let\@oddfoot\@empty
  \let\@evenfoot\@oddfoot
}
\makeatother
 
\usepackage{amssymb,amsmath,mathtools}

\usepackage{graphicx}
\usepackage{cancel}
\usepackage[ruled,vlined]{algorithm2e}
\usepackage{hyperref}
\usepackage[dvipsnames]{xcolor}

\usepackage[colorinlistoftodos,textwidth=4cm,shadow]{todonotes}

\usepackage{setspace}
\newcounter{Igor}

\date{}
\title {\textbf{How Fast are Elastic Domino Waves?}}

\author{Daniel Ding }
\author{Clement Lau }
\author{Jorrit Westerhof}
\author{Lotte van der Hoeven}
\author{Lieke Kampstra}
\author{Patrick van der Beek}

\author{Igor Ostanin \footnote{Corresponding author, e-mail:i.ostanin@utwente.nl.}}

\address{Multi-Scale Mechanics (MSM), Faculty of Engineering Technology, MESA+, University of Twente, P.O. Box 217, 7500 AE Enschede, The Netherlands.}

\begin{document}

\begin{abstract}
The paper is concerned with the problem of toppling propagation velocity in elastic, domino-like mechanical systems. We build on the work of Efthimiou and Johnson, who developed the theory of perfectly elastic collisions of thin rigid dominoes on a frictional foundation. This theory has been criticised for the lack of correspondence with the experimental observations, in particular, prediction of infinite propagation velocity for zero spacing between dominoes, as well as the inability to represent the collective nature of collisions in real domino systems. In our work we consider a more realistic scenario of dominoes of finite stiffness and obtain a theory of fast elastic domino waves, taking into account a limit velocity of the perturbation propagation in the system of dominoes. Moreover, finite collision time allows to extract dynamic quantities of collisions and establish upper and lower borders for domino separations where the theory could still be applied. Our discrete element simulations support our theoretical findings and shed light on the nature of collective interactions in the nearly-elastic domino chains.
\end{abstract}

\maketitle

\section{Introduction}

The phenomenon of domino wave -- the propagation of toppling in equal-sized and equispaced rectangular blocks(Fig. 1(A)), that we will further refer to as dominoes -- attracted attention of many researchers. Probably the first call to consider the mechanics behind the domino waves belongs to Daykin (1971) \cite{Daykin_1971}, who suggested to propose a set of reasonable assumptions to solve the problem of the velocity of domino wave with mathematical rigor. McLachlan et al \cite{McLachlan_1983} used the analysis of problem dimensions to establish the scaling law for wave propagation velocity -- it was shown that it has the shape

\begin{equation} \label{eq1}
	v = \sqrt{gl}G\left(\frac{d}{l}\right),
\end{equation}

where $g$ is the acceleration of free fall, $l$ is the length of the domino, $d$ is the spacing between dominoes and $G$ is some unknown function. The work \cite{McLachlan_1983} also demonstrated that the experiments confirm the suggested scaling. Subsequently, somewhat more refined scaling laws were suggested (e.g. \cite{Bohua_2020, Bohua_2021}) that explored the role of domino thickness, path curvature etc.

The work of Bert \cite{Bert_1986} have presented a complete solution for the dynamics of colliding dominoes in terms of elliptic integrals. Stronge \cite{Stronge_1987} for the first time demonstrated the existence of analytical limit velocity of domino wave propagation that does not depend on the initial perturbation and associated it with the presence of friction between colliding dominoes.  
 
The question from Dutch national science quiz of 2003 motivated the work of van Leeuwen \cite{vanLeeuwen_2010}, which was published as a pre-print nearly concurrently with the paper by Efthimiou and Johnson \cite{Johnson_2007} (hereafter - EJ). These two works, done independently from earlier works \cite{Bert_1986, Stronge_1987} sparkled a new wave of interest to the problem of domino collisions. Following earlier work of Shaw \cite{Shaw_1978}, van Leeuwen considers the collisions between dominoes as inelastic, which leads him to the concept of ``domino trains'', or solitons. In contrast, EJ model \cite{Johnson_2007}, as well as the earlier work \cite{Bert_1986}, considers propagation of domino wave as the series of pair-wise elastic collisions. Obviously, both approaches are valid in a certain range of domino properties. Intermediate regimes between two theories are complex (see the discussion in Subsection 3.2 below), it is therefore hard to derive one as the limit case of another. Clearly, the assumptions of \cite{vanLeeuwen_2010} are a lot closer to reality for regular domino systems, a pair collision regime of \cite{Johnson_2007} is hard to achieve in experiment.

These two works inspired a large number of subsequent developments on the problem, exploring its different aspects (see, e.g.,  the recent works \cite{Shi_2019, Bohua_2020, Bohua_2021, Cantor_2022, Pola_2023}). However, surprisingly, there were no attempts to look deeper into another important feature of the problem - contact interactions between dominoes. Starting from the scaling law (\ref{eq1}), it was assumed that the domino propagation velocity should be independent on the density and the Young's modulus of domino's material. This assumption is justified when the collision time is negligibly small compared to the free fall time of the domino. However, this assumption breaks for small separations between dominoes, which results in a singular behavior of domino wave propagation velocity according to \cite{Johnson_2007}. Of course, no mechanical interactions between dominoes can propagate faster than the speed of P-wave in the domino's material. 

This deficiency of EJ theory was pointed out in \cite{Larham_2008} -  it was noted that the theoretical predictions diverge significantly from the experimental observations, especially in the region of small separations between dominoes.

 Another important consequence of the assumption of ``instant'' elastic collisions are presumed infinite contact forces and angular moments acting on dominoes at the moment of collision (this question is revisited below in the Subsection 2.4). This explicitly contradicts the assumption of the ``sufficient'' force of friction to exclude slip at the domino's support point, since ``sufficient'' in this case implies the infinite coefficient of friction, or the hinge-like connection of the domino with the foundation. This makes it hard to relate the predictions of theory with the real domino-like systems.

These considerations led to the idea of fundamental deficiency of EJ model and it was largely labeled by the community as irrelevant and insufficient for adequate description of domino systems. 

However, the model itself is not internally contradictory -- the pair-collision regimes are certainly achievable, and the phenomenon of finite-velocity domino wave, can indeed exist even in a perfectly conservative system, as predicted by EJ theory.

In our work, we attempt to build on the assumptions of EJ theory and incorporate the finite contact stiffness between dominoes. It appears that this modification leads to a more complex scaling of domino wave velocity than the one predicted by mcLachlan. In our model, the domino wave does not exceed the wave velocity in the chain of dominoes in elastic contact. 

Introduction of finite collision time between perfectly elastic dominoes allowed us to quantify the dynamics of the collision and establish the bounds on the parameters of the system that admit EJ-like behavior. It was demonstrated both analytically and numerically that pair-wise collisions between dominoes are strictly attributed to ideally conservative system, whereas collective collisions (domino trains) are associated with the presence of energy dissipation.

Our theoretical prediction of domino wave velocity reasonably agrees with the results of Discrete Element Method (DEM) numerical modeling.

The paper is organized as follows. In the next section, we describe our analytical model and highlight some of its properties. The third section compares the predictions of our model with the results of experiments and DEM numerical simulations, and establishes the borders of applicability of the modified theory. The concluding section summarizes and discusses our findings.

\section{Finite collision time domino wave theory}

\subsection{The case of perfectly rigid dominoes}

Let us first concisely overview the major results of the model presented by Efthimiou and Johnson \cite{Johnson_2007} that we have chosen as the baseline for our derivations. They modeled a row of dominoes as a system of initially vertical, infinitely thin rigid rods of height $l$, equispaced at distance $d$ apart, standing on frictional horizontal foundation (Fig. 1(B)). The inertial properties of a domino are represented with a point mass $m$ at the upper tip of the massless rod.\footnote{Other mass distributions can be straightforwardly considered (see, e.g. \cite{Bert_1986}), a simple concentrated mass case is chosen in \cite{Johnson_2007} for brevity of analytical expressions.} The friction between dominoes and the foundation was assumed to be sufficient to exclude slip; therefore, the domino toppling is viewed as a purely rotational motion. The friction between dominoes is neglected. The following notations were used (our notations mostly follow \cite{Johnson_2007}):

\begin{itemize}
	\item $\Omega_n$ is the angular velocity of the domino $n$ immediately after collision with the domino $n-1$
	
	\item $\Omega_{fn}$ is the angular velocity of the domino $n$ immediately before collision with the domino $n+1$
	
	\item $\Omega_{bn}$ is the angular velocity of the domino $n$ immediately after collision with the domino $n+1$
	
	\item $A_{n}$ is the point of support (center of rotation) of domino $n$
	
	\item $\beta = \arcsin(d/l)$ is the angle of inclination of the domino $n$ at the moment of collision with the domino $n+1$ (Fig. 1(C))
	
	\item $I = m l^2$ is the moment of inertia of the domino $n$ with respect to $A_{n}$
	
\end{itemize}

\begin{figure}
	\begin{center}
		\includegraphics[width=12.5cm]{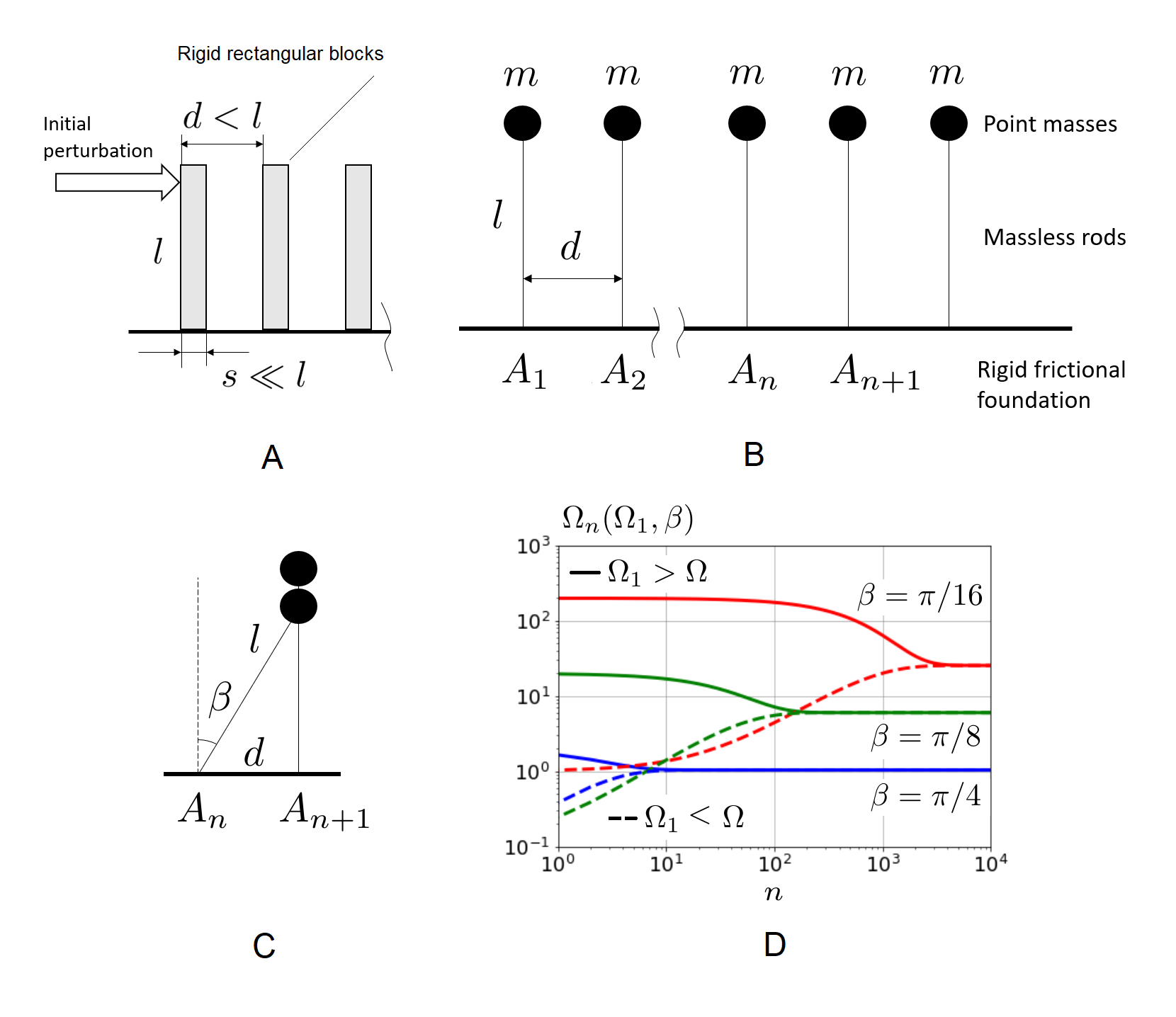}
		\protect\caption{(A) A system under consideration. (B,C) The suggested simplified model: the configuration of dominoes at the initial moment (B) and the moment of collision of $n$-th and $n+1$-th dominoes (C). (D) Angular velocity of domino $\Omega_n$ as the function of its number $n$ for three different angles $\beta$, each for two initial angular velocities $\Omega_1<\Omega$ and $\Omega_1>\Omega$, where $\Omega$ is the limit angular velocity given by \ref{eq9}, ($\sqrt{g/l}=1$).}
	\end{center}
\end{figure}

Assuming precise conservation of energy and angular momentum about the point $A_{n+1}$ during collision, we can write down the following system of equations, linking $\Omega_n$, $\Omega_{fn}$, $\Omega_{bn}$.

\begin{equation} \label{eq2}
	\begin{cases}
		\frac{1}{2}I\Omega_{fn}^2 = \frac{1}{2}I\Omega_{bn}^2 + \frac{1}{2}I\Omega_{n+1}^2\\
		\Omega_{fn}\cos^2{\beta} = \Omega_{bn}\cos^2{\beta} + \Omega_{n+1}  
	\end{cases}
\end{equation}

It's solution can be given as:

\begin{equation} \label{eq3}
	\begin{split}
		\Omega_{n+1} & = 
		f_{+}\Omega_{fn} \\
		\Omega_{bn} & = \frac{\Omega_{n+1}}{ f_{-}}  
	\end{split}
\end{equation}

where

\begin{equation} \label{eq4}
 f_{\pm} = \frac{2}{\cos^2 \beta \pm 1/\cos^2\beta}  
\end{equation}

Further, considering energy balance for the domino falling in a gravity field, one can write:

\begin{equation} \label{eq5}
\frac{1}{2}I\Omega_{n}^2 + mgl = \frac{1}{2}I\Omega_{fn}^2 + mgl \cos \beta  
\end{equation}

Combining (\ref{eq3}) and (\ref{eq5}) we have 

\begin{equation} \label{eq6}
	\Omega_{n+1}^2 = f_{+}^2 \Omega_{n}^2 + b
\end{equation}

where    

\begin{equation} \label{eq7}
	b = \frac{2g}{l}f_{+}^2(1-cos \beta)
\end{equation}

Solving \cite{Johnson_2007} for the $n$-th term of the mixed progression (\ref{eq6}) we have

\begin{equation} \label{eq8}
	\Omega_{n}^2 = f_{+}^{2(n-1)} \Omega_{1}^2 + b \frac{1-f_{+}^{2(n-1)}}{1-f_{+}^{2}}
\end{equation}

The nice property of this solution \cite{Johnson_2007} is that it predicts that the limit angular velocity at $n \rightarrow \infty$ does not depend on the angular velocity $\Omega_1$ caused by initial external push:

\begin{equation} \label{eq9}
	\Omega = \lim_{n \rightarrow \infty}  \Omega_{n}^2 = \frac{2g}{l}(1-\cos \beta) \frac{f_{+}^2}{1-  f_{+}^2} 
\end{equation}
Fig. 1(D) qualitatively demonstrates some properties of the solution \ref{eq8}. We can see that for large enough $n$ it converges to limit velocity $\Omega$, which occurs irrespectively of the initial angular velocity $\Omega_1$ and angle $\beta$, although the speed of convergence dramatically increases with increase of $\beta$ and occurs somewhat faster for the case of increasing angular velocity($\Omega_1<\Omega$). Unlike stated in \cite{Stronge_1987}, the property of stable limit velocity exists even in the case of fully conservative system and is not directly related to presence of dissipation/friction.  
  
It is easy to demonstrate that for $n \rightarrow \infty$ we have

\begin{equation}  \label{eq9a}
	\begin{split}
		\Omega_{n}^2 &= \Omega_{n+1}^2 \\
		\frac{I \Omega_{bn}^2}{2} &= mgl (1-\cos{\beta})
	\end{split}
\end{equation}

\textit{i.e.}, the amount of potential energy released by the fall of the domino $n$ in the gravity field precisely equals to the kinetic energy of domino $n$ bouncing back after the collision. This means that, irrespectively of the angle $\beta$, the domino $n$ will reach its original vertical position at zero angular velocity after collision with the domino $n+1$, while the latter will topple with precisely the same angular velocity profile as the domino $n$. Thus, we can see that the accumulation of released potential energy does not occur, and stable limit propagation velocity exists in a fully conservative system. The picture, however, changes significantly in presence of energy dissipation - the domino $n$ that bounced back after the collision is not stabilized, which subsequently causes the secondary (collective) wave, gradually overtaking the initial (pair-collision) wave -- please see the discussion in Section 3 and the corresponding video demonstrations in the Supplementary Information.

The speed of the domino wave is defined by the time period between two subsequent domino collisions. This time period can be derived from the integration of the equation of energy balance for the domino inclination to arbitrary angle $\theta \in (0, \beta)$,

\begin{equation} \label{eq10}
	\frac{1}{2}I\Omega_{n}^2 + mgl = \frac{1}{2}I \left({\frac{d\theta}{dt}}\right)^2 + mgl \cos \theta  
\end{equation}

or 

\begin{equation} \label{eq11}
	\int_{0}^{T_{n}}dt = 
	\int_{0}^{\beta} \frac{d \theta}{\sqrt{\Omega_{n}^2+ \frac{2g}{l} - \frac{2g}{l} \cos \theta}}  
\end{equation}

The closed form solution is given in terms of elliptic integrals of the first kind:

\begin{equation} \label{eq12}
	T_{n} = \frac{2}{\sqrt{a_n+c}}\left[ K(\kappa_n) - F\left( \frac{\pi - \beta}{2}, \kappa_n \right) \right]   
\end{equation}

where $a_n = \Omega_n^2+2g/l$, $c = 2g/l$, and $\kappa_n = \sqrt{\frac{2c}{a_n+c}}$; here the elliptic integrals are defined as 

\begin{equation} \label{eq13_1}
	\begin{split}
		F(\alpha, k) &= \int_{0}^{\alpha} \frac{d \alpha} { \sqrt{1-k^2 \sin^2{\alpha}}} ( 0 \leq k < 1),\\
		K(k) &= F(\pi/2, k)   
	\end{split}
\end{equation}

In the limit of large $n$, the time $T_n$ approaches a limiting value

\begin{equation} \label{eq13}
	T = \frac{2}{\sqrt{a+c}}\left[ K(\kappa) - F\left( \frac{\pi - \beta}{2}, \kappa \right) \right]   
\end{equation}

where $a = \Omega^2+2g/l$,  $\kappa = \sqrt{\frac{2c}{a+c}}$.

This result straightforwardly leads to the expression for the domino wave velocity:

\begin{equation} \label{eq14}
	v = \frac{d}{T} =
	\frac{\sqrt{a+c}}{K(\kappa)-F\left( \frac{\pi - \beta}{2}, \kappa \right)} 
\end{equation}

This expression can be explicitly re-written in McLachlan's form: $v = \sqrt{gl}G(\frac{d}{l})$ (see \cite{Johnson_2007} for the closed-form expression for $G$).

\subsection{The case of compliant dominoes}

The EJ theory, presented above, leads to a clearly non-physical phenomenon for small separations between the dominoes -  the theory predicts singular behavior of the velocity with the scaling $1/\beta$, or $l/d$. 

To address this problem, we generalize the EJ model by taking into account the finite collision time between dominoes. Our analysis is based on the following assumptions:

\begin{itemize}
	
	\item Interactions between dominoes remain perfectly elastic, but the contact stiffness is not anymore infinite. This leads to finite overlaps between dominoes and  finite collision time, comparable with the limit time of the domino's free fall, given by (\ref{eq13}) - see the diagram on Fig. 2.
	
	\item The collision between dominoes is assumed to be represented by the unconstrained head-on collision of two equal spherical particles in translational motion. Particles are assumed to have the finite radius $R$, which is considered to be much less than $l$ and $d$, and therefore does not affect the angle $\beta$. However, the maximum overlap between particles $\delta$ is assumed to be much less than $R$. 
	
	\item The contact stiffness is constant (the assumption known to DEM community as the ``linear'' contact model \cite{Cundall_1979}). This stiffness is defined based on the radius of the particles in contact and their Young's modulus (see below).
	
	\item The collision time is defined as the half-period of free vibration of the equivalent undamped spring-mass system, meaning that the work of external force (gravity) is neglected during the collision.   
\end{itemize}

\begin{figure}
	\begin{center}
		\includegraphics[width=10.0cm]{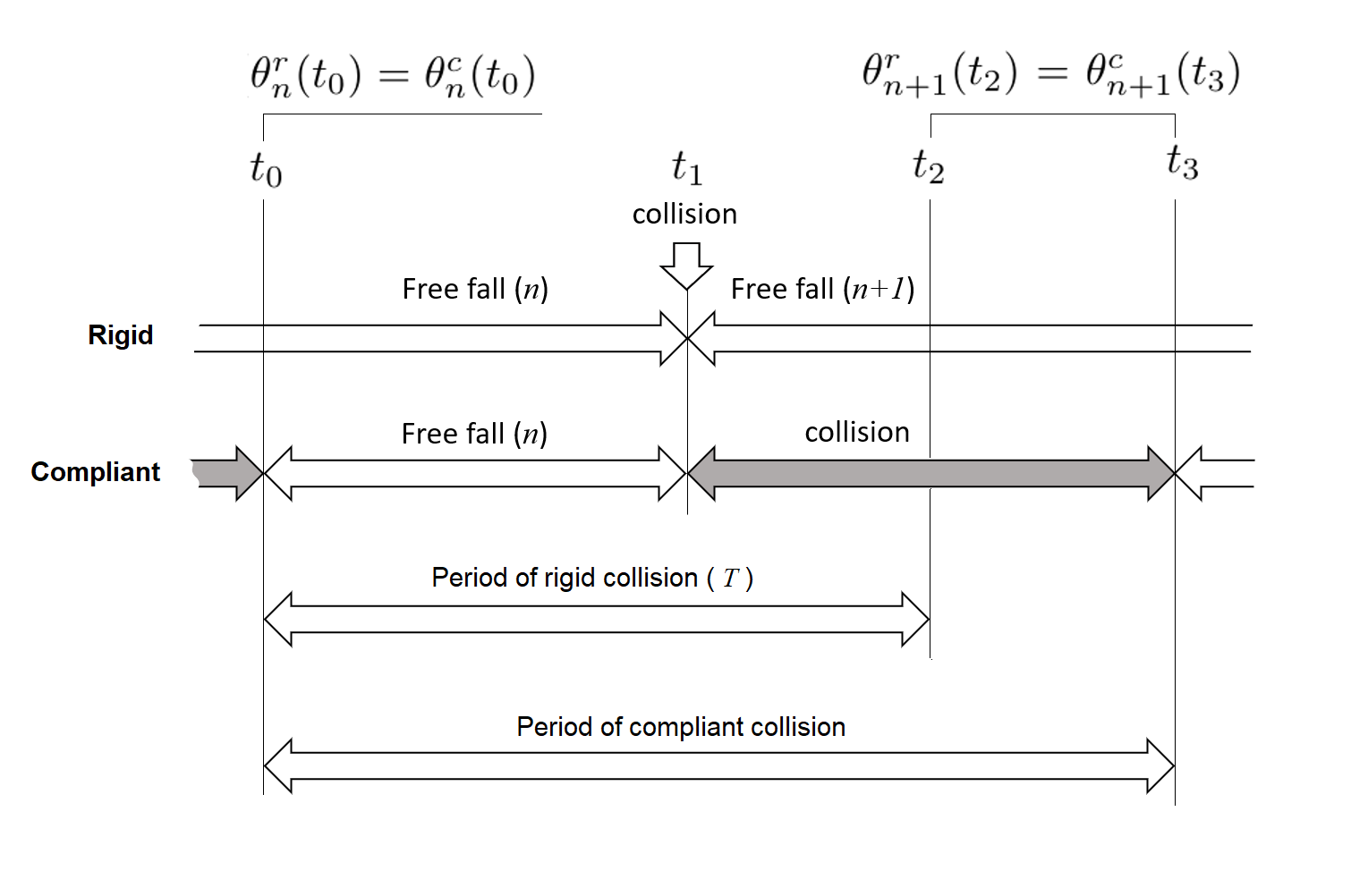}
		\protect\caption{Comparative time diagram of one period of rigid (instantaneous) and compliant (finite time) collision}.
	\end{center}
\end{figure}

In order to establish wave propagation velocity in the system of compliant dominoes, let us have a closer look at the diagram in Fig. 2, comparing the motion in the system of rigid and compliant dominoes. Consider the situation when rigid and compliant chains with otherwise identical parameters are synchronized at the initial moment $t_0$, corresponding to the moment of detachment of compliant dominoes: 

\begin{equation} \label{eq15}
	\theta^r_n(t_0) = \theta^c_n(t_0) 
\end{equation}

The motion in both chains is identical till the moment $t_1$, corresponding to the instance of collision in the rigid chain (or the beginning of collision process in compliant chain). Let us denote the end of collision in the compliant chain as $t_3$, and the moment $t_2$ such that

\begin{equation} \label{eq16}
	\theta^r_{n+1}(t_2) = \theta^c_{n+1}(t_3) 
\end{equation}

Easy to see that the period between sequential collisions in the rigid system $T^{rig} = t_2 - t_0$ differs from the similar period in the compliant system $T^{compl} = t_3 - t_0$ by the term $\Delta t = t_3 - t_2$. 

Assuming constant stiffness collision and harmonic acceleration, one can establish relation between $\Delta t$ and the collision time $T_c = t_3 - t_1$. The position reached by the compliant domino $n+1$ at the moment of detachment can be expressed in this case as

\begin{equation} \label{eq17}
	 \theta^c_{n+1}(t_3) = \theta^c_{n+1}(t_1) + \frac{\Omega_{n+1}}{2} \int_{0}^{T_c} \left(1 - \cos{\frac{\pi}{T_c} t}\right)dt = \Omega_{n+1} \frac{T_c}{2}
\end{equation}
 
The same position is expressed via angular velocity of rigid domino as 

\begin{equation} \label{eq18}
	\theta^r_{n+1}(t_2) = \theta^r_{n+1}(t_1) + \Omega_{n+1} \Delta t
\end{equation}

Given that $\theta^r_{n+1}(t_1) = \theta^c_{n+1}(t_1)$ we immediately get

\begin{equation} \label{eq19}
	\Delta t = \frac{T_c}{2}
\end{equation}
The domino wave velocity in the system of compliant dominoes can therefore be written as 
  
\begin{equation} \label{eq20}
	v = \frac{d}{T + T_c/2} 
\end{equation}

where $T$ is given by (\ref{eq13}). 
Based on the assumptions listed above, we can express the collision time as: 

\begin{equation} \label{eq21}
	T_c = \pi \sqrt{\frac{m_{r}}{k_{r}}} 
\end{equation}

here $m_{r}, k_{r}$ are equivalent mass and stiffness of spring-mass system representing the collision of dominoes $n$ and $n+1$. Note that $T_c$ does not depend on $n$, therefore, the corresponding limits are omitted here and below. 

Let us have a closer look at the parameters $m_{r}, k_{r}$.

In case of head-on collision of two unconstrained particles with masses $m_1,m_2$ and contact stiffnesses $k_1, k_2$ we can write down:

\begin{equation}  \label{eq22}
	\begin{split}
		m_{r} = \frac{m_1m_2}{m_1+m_2} \\
		k_{r} = \frac{k_1k_2}{k_1+k_2}
	\end{split}
\end{equation}

Stiffnesses $k_1, k_2$ are defined as 

\begin{equation}  \label{eq23}
	\begin{split}
		k_1 = \frac{E_1 \pi R^2}{R} = \pi R E_1 \\
		k_2 = \frac{E_2 \pi R^2}{R} = \pi R E_2,
	\end{split}
\end{equation}

given $E_1 = E_2 = E$, the contact stiffness is given by 

\begin{equation}  \label{eq24}
	k_{r} = \frac{\pi R E}{2} 
\end{equation}

Assuming that the rotations are negligibly small during collision, we can consider collision in terms of translational dynamics of concentrated masses (Fig. 3(A)). The equivalent mass of the first domino is $m$. In order to ensure the conservation of the moment of inertia $I = ml^2$ of the domino $n+1$ with respect to its foundation $A_{n+1}$, we need to assume the collision with the concentrated mass $m'$ (Fig. 3(B)), such that

\begin{equation} \label{eq25}
	I = m l^2 = m' s^2 = m l^2 \cos^2 \beta 
\end{equation}

The equivalent mass of the second domino is therefore 
 
\begin{equation} \label{eq26}
	m' = \frac{m}{\cos^2 \beta}
\end{equation}

\begin{figure}
	\begin{center}
		\includegraphics[width=10cm]{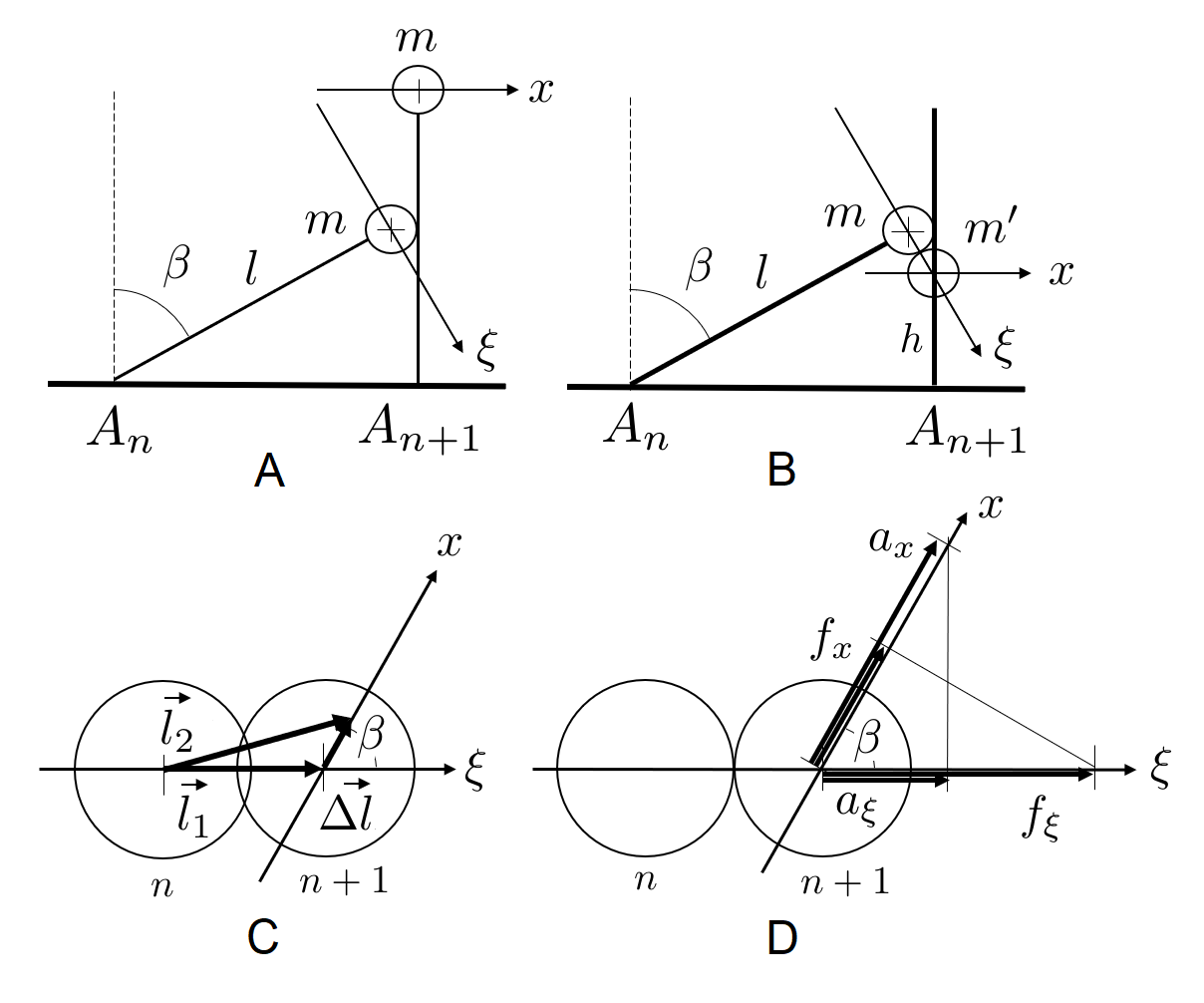}
		\protect\caption{Derivation of the collision time. (A) Initial problem (B) representation as a head-on constrained collision of two point masses. (C,D) Illustrations on the derivations of effective stiffness (C)  and mass (D) of unconstrained collision.}
	\end{center}
\end{figure} 
 
We now have the head-on soft collision of two concentrated masses However, both masses are constrained - the mass $n$ can move only along $\xi$, whereas the mass $n+1$ moves only along $x$. Such constraints are usually called holonomic (geometric and integrable) in theoretical mechanics. Computing the collision time requires integration of the motion of the system shown in Fig. 3(B), which, in general case is only possible numerically. However, under the assumptions discussed above, we can estimate the collision time based on similarity with head-on (1D) collision in a system of two unconstrained masses. Below we derive the equivalent mass and stiffness of such a system.     
 
Consider the constrained system of two masses shown in Fig. 3(B). The mass $n$ hits the initially resting mass $n+1$ and bounces back along $\xi$ axis, while the mass $n+1$ starts to move along the $x$ axis. Consider two distinct moments of time during the collision: $t_1$, when the two particles are head-on at the distance $l_1$, and $t_2$, when the particle $N+1$ displaced slightly along $x$ relative to initial head-on position. The vector difference between $\vec{l_1}$ and $\vec{l_2}$ is $\vec{\Delta l}$.  As discussed above, overlaps between particles are considered to be much smaller than the particle radius. Easy to see (Fig. 3(C)) that the change of the intercenter distance between two particles in this case is:

\begin{equation} \label{eq27}
	\begin{split}
		l_2-l_1 = \sqrt{(l_1 + \Delta l \cos \beta)^2 + (\Delta l \sin \beta)^2} = \sqrt{l_1^2 + 2 l_1 \Delta l \cos \beta + \Delta l^2 \cos^2 \beta + \Delta l^2 \sin^2 \beta} \\ 
		= l_1 \sqrt{1 + \frac {2 \Delta l}{l_1} \cos \beta + O \left( \left( \frac {\Delta l}{l_1}\right)^2\right)}
		\approxeq l_1 \left( 1 + \frac{\Delta l}{l_1} \cos \beta \right)
		=l_1 + \Delta \xi  
	\end{split}
\end{equation}

Therefore, we can see that if $\Delta l \ll R$, then up to the leading terms, the effective contact stiffness along $\xi$, as well as the time instances of contact formation and breakage (and therefore, the collision time) will not be perturbed by the motion transversal to $\xi$. We can therefore conclude that the stiffness along $\xi$ should be considered the same as in the case of unconstrained motion.

Let us then have a look at the effect of constraint on the dynamics along $\xi$. Easy to see (Fig. 3(D)) that due to the presence of constraint, spring force $f_{\xi}$ can only cause the acceleration along $x$ ($a_x$), causing, in turn, the projected acceleration along $\xi$:

\begin{equation} \label{eq28}
	a_{\xi} = a_{x} \cos \beta = \frac{f_{x}}{m'} \cos \beta =
	\frac{f_{\xi} \cos \beta}{m'} \cos \beta =
	\frac{f_{\xi}} {m''}, \\
	\end{equation}        

where

\begin{equation} \label{eq29}
	m'' = \frac{m'}{\cos^2 \beta} = \frac{m}{\cos^4 \beta}
\end{equation}        

Note that in case of $\beta = 0$ the effective mass $m''$ exactly coincides with $m$, while in case of $\beta = \frac{\pi}{2}$ we have $m'' \rightarrow \infty$. 

Therefore, under aforementioned assumptions, the collision between two dominoes can be viewed as a head-on collision of two unconstrained masses $m_1 = m$ and $m_2 = m''$. A pairwise interaction of dominoes can thus be reduced to a single spring-mass system with the parameters   
   
\begin{equation} \label{eq30}
	\begin{split}
		k_{r} &= \frac{\pi R E}{2}, \\
		m_{r} &= \frac{m_1 m_2''}{m_1 + m_2''} = 
		\frac{m}{1+\cos^4 \beta}
	\end{split}
\end{equation}

Therefore we can express the collision time as 

\begin{equation} \label{eq31}
	T_c(k,m,\beta) = \pi \sqrt{ \frac{ 2 m }{k (1+\cos^4 \beta)} }
\end{equation}

or, in terms of the particle radius $R$, mass $m$ and the material Young's modulus $E$:

\begin{equation} \label{eq32}
	T_c(R,E,m,\beta) = \sqrt{ \frac{2 \pi m }{R E (1+\cos^4 \beta)} }
\end{equation}

\begin{figure}
	\begin{center}
		\includegraphics[width=12.5cm]{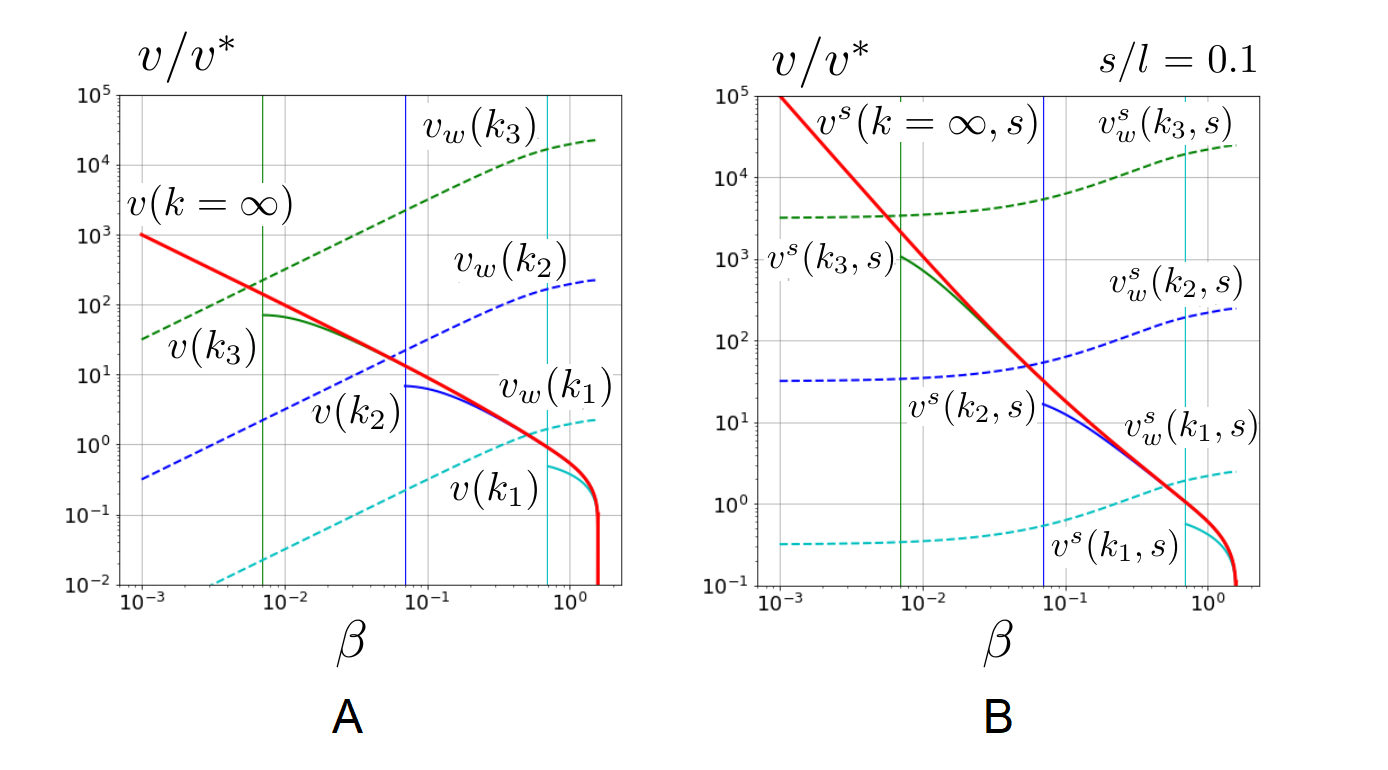}
		\protect\caption{Velocities of domino wave propagation as functions of angle $\beta$, for infinitely thin (A, $v(k)$) and finite thickness (B, $v^s(k, s)$) dominoes. The plots are given for infinitely stiff dominoes (red lines) and three different values of stiffness. The plots are compared with the corresponding P-wave velocities.}
	\end{center}	
\end{figure}

One can see that due to the finite quantity $T_c/2$ in the denominator of (\ref{eq20}), the wave propagation velocity can not be singular for small separations. 
It is important to note that the expression (\ref{eq20}) is applicable only until the collision events are distinct, meaning that the collision of dominoes $N$ and $N+1$ is not initiated before the complete detachment of dominoes $N-1$ and $N$. It is easy to demonstrate that this condition can be re-written as $T>T_{c}/2$. Otherwise, the equations of harmonic pair-wise collisions are not valid anymore. For $0<T<T_{c}/2$, one can expect complex patterns associated with the emergence and disappearance of interactions, while for $T = 0$ we end up with classical equations for wave propagation in a dispersive spring-mass chain, with velocity of propagation dependent on the frequency of the initial perturbation. 

Fig. 4 (A) illustrates the dependence of the domino wave velocity on the angle $\beta$. The plots are given for three different values of stiffness: $k_1 = k^*$, $k_2 = 10^4 k^*$, $k_3 = 10^8 k^*$, where $k^*$ is the value of stiffness defined by $ l/g = m/k^*$ for a given mass $m$. The curve $v(k = \infty)$ is provided for the reference.

The domino wave velocities are compared with acoustic P-wave velocity in a spring-mass chain with the masses $m_r$, stiffnesses $k_r$ and spring length $d$:

\begin{equation} \label{eq33}
	v_w = d \sqrt{\frac{k_r}{m_r}}
\end{equation}

 One can see that the domino wave predicted by our theory can not be faster than the corresponding P-wave velocity in the system.

\subsection{Dominoes of finite thickness}

It is useful to consider the generalization of our expressions for the dominos of finite thickness $s$. Here and below $s/l$ is considered to be small. The effect of finite thickness of a domino is two-fold. First, finite thickness creates the potential energy minimum that leads to domino's vertical stability in a certain range of inclination angles. This change in the potential energy relief also affects the integral (\ref{eq11}) defining the free fall time. For the purpose of wave velocity estimation, we neglect this effect as quadratic with respect to $s/l$. Second, domino's finite thickness effectively increases the velocity:
  
\begin{equation} \label{eq34}
	v^s = \frac{d+s}{T+T_c/2}
\end{equation}

When comparing our results with the experiments and DEM simulations, we use thickness-adjusted expression for velocity (\ref{eq34}). The wave velocity in such structure is evaluated as   
 
\begin{equation} \label{eq35}
	v_w^s = (d+s) \sqrt{\frac{k_r}{m_r}}
\end{equation}

Fig. 4(B) illustrates the effect of finite domino thickness ($s/l = 0.1$) on the curves that are depicted in Fig. 4(A).  

\subsection{Range of applicability of the theory}

DEM simulations that are discussed below, feature more complex mechanical behavior than the one predicted by the theory above. Assuming validity of assumptions of 2D motion of dominoes and their sequential pair-wise interactions, the key factor defining the applicability of the extended EJ theory is the friction between domino and foundation. If the friction is sufficient, the domino rotates around the support point, if not - the point of support can slide along the foundation. Fig. 5(A) illustrates possible types of domino motion that can be initiated in this case. In the analysis below we establish the bounds within which sliding of the domino's foundation is impossible.

It is convenient to give the bounds in terms of magnitude of the contact force, emerging during harmonic collision.

\begin{equation} \label{eq36}
	\int_{0}^{T_c} F_x dt = \cos{\beta} \int_{0}^{T_c} F_m \sin{\left( \frac{\pi}{T_c}t \right)}dt = m'v = m' \Omega l \cos{\beta}  
\end{equation} 

This leads to

\begin{equation} \label{eq37}
	F_m = \frac{\pi m \Omega l}{2 T_c \cos^2 \beta}  
\end{equation} 

where $\Omega$ is given by (\ref{eq9}), and $T_c$ is defined according to (\ref{eq32}). One can note that instant collision ($T_c = 0$) leads to infinite contact force $F_m$.

The domino can exhibit initiation of forward or backward rotation when the moment created by the contact force at collision point is not compensated by the moments produced by the frictional force and (in case of domino's finite thickness) reaction of the support to the gravitational force. This condition can be written as:

\begin{equation} \label{eq38}
	\frac{F_m}{m g} > \frac{\mu+s/l}{\cos{\beta} \left|2 \cos{\beta}-1\right|}  
\end{equation} 
 
Note that this shape is valid for both the collisions above and below the level of the domino's center of mass ($\beta = \pi/3$). 

Straightforward considerations allow to conclude that the translational sliding of the domino along the foundation can be initiated in one of the two following cases:

\begin{itemize}

\item Case 1:

1) Horizontal projection of $F_m$ exceeds the frictional force $\mu m g$, which leads to:
 
 \begin{equation} \label{eq39}
 	\frac{F_m}{ m g} > \frac{\mu}{\cos{\beta}}  
 \end{equation}

2) The collision is below the level of resting domino's center of mass ($\beta>\pi/3$).

3) The moment acting on the domino exceeds one exerted by frictional force but is lower than the sum of moments exerted by frictional and gravitational force:

\begin{equation} \label{eq41}
	\frac{\mu}{\cos{\beta}(1 - 2 \cos{\beta})}<
	\frac{F_m}{m g} < \frac{\mu+s/l}{\cos{\beta}(1 - 2 \cos{\beta})}  
\end{equation}

\item Case 2:

1) Condition (\ref{eq39}) is met.
2) The toppling moment exerted by contact force does not exceed the stabilizing moment exerted by the gravity: 
   
\begin{equation} \label{eq42}
	\frac{F_m}{m g} < \frac{s}{2 l \cos{\beta}}  
\end{equation}

\end{itemize}

\begin{figure}
	\begin{center}
		\includegraphics[width=12.5cm]{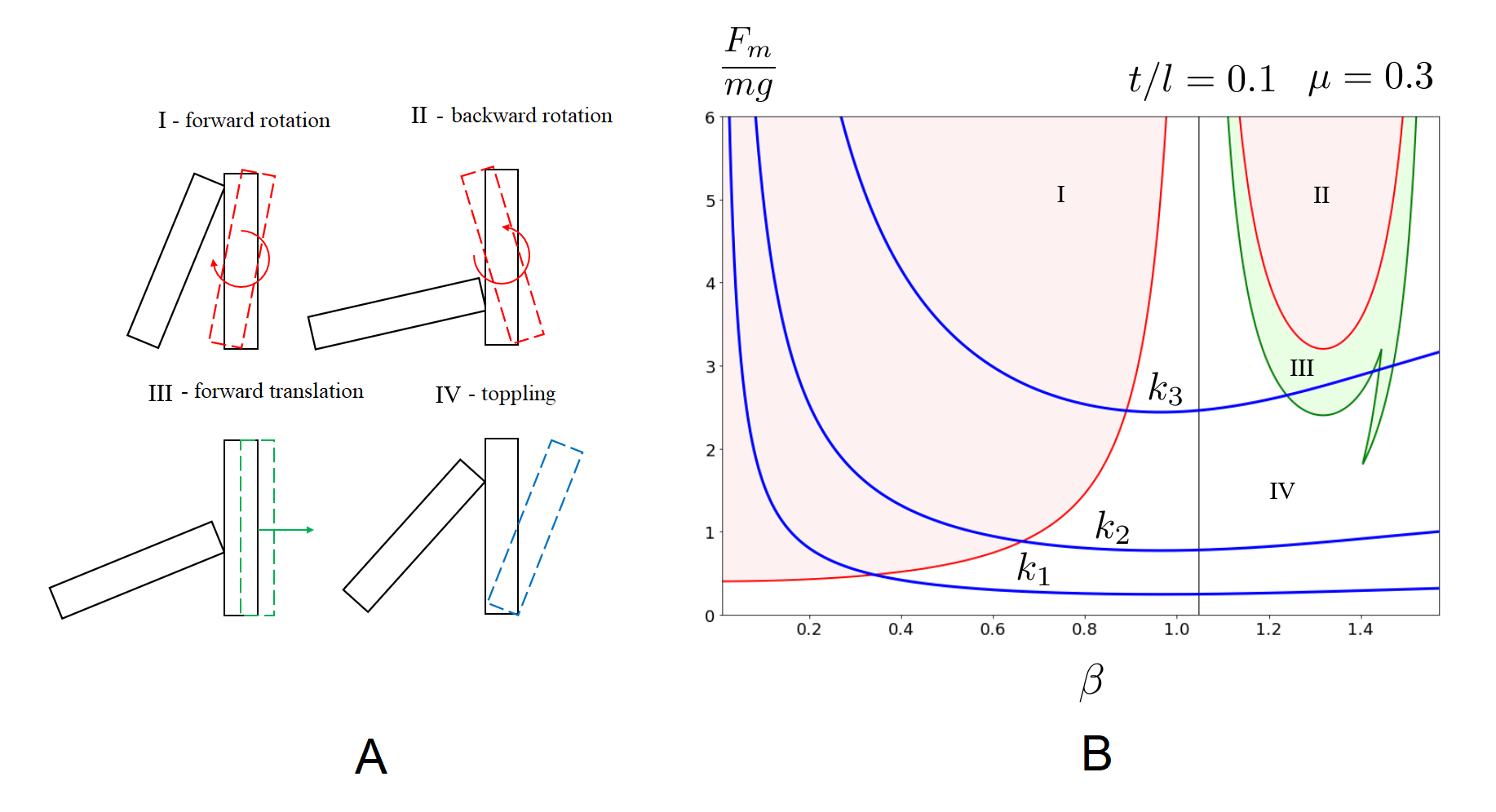}
		\protect\caption{(A) Possible mechanisms of domino sliding, (B) Diagram detailing the boundaries between these mechanisms. Blue lines indicate magnitudes of contact forces, given by (\ref{eq37}), for few different contact stiffnesses ($k_1 = k^*$, $k_2 = 10 k^*$, $k_3 = 100 k^*$, $k^*$ is defined above). The plots are given for $s/l = 0.1$, $\mu = 0.3$.}
	\end{center}	
\end{figure}

Fig. 5(B) gives ranges of angles where scenarios I-IV, depicted in Fig. 5(A), can occur. It is worth noting that the expressions (\ref{eq38}-\ref{eq42}) give the ``tightest bounds'', beyond which the theoretical assumption of resting foundation does not hold. The further domino dynamics after onset of sliding is hard to establish within the analytical model. The numerical modeling provided below gives some idea on further dynamic evolution of the domino chain in these cases.

\section{DEM modeling of dominoes}

We used the discrete element method \cite{Cundall_1979} to study the domino effect, employing open source DEM package YADE \cite{Yade_2021} in the calculations \footnote{For the verification purposes, our simulation framework was also implemented in the open-source package MercuryDPM \cite{weinhart2020fast, Ostanin_2023}, simulations in both frameworks showed good agreement.}. The dynamics of equal-sized rigid spherical beads with mass $m_p$, radius $R_p$, volume $V_p = \frac{4}{3} \pi R^3$ and moment of inertia $\frac{2}{5}m_pR^2$ was computed using the velocity Verlet time integration scheme. The domino parts were modeled as rigid assemblies (clumps) of the beads (Fig. 6(A)). Rigid clumps are widely used to model nonspherical particles in DEM. Compared to the alternative approaches \cite{Podlozhnyuk_2017, Govender_2016} often used in DEM, they offer high performance, combined with a wide library of contact models -- any model that is available for a spherical particle is also immediately available for a rigid clump. In our case we use linear contact model, allowing to directly relate our theoretical analysis with DEM simulations. In presence of non-zero friction between the beads, classical Cundall and Strack no-slip contact model \cite{Cundall_1979} was employed. Two separate friction coefficients were specified - $\phi_d = 0$ is associated with all the clumped spheres, $\phi_f$ (unless otherwise noted, $\phi_f = 0.3$) is used for the bottom layer of spheres and the frictional foundation. For every pair contact between entities $1$ and $2$, $\phi = \min{}{(\phi_1, \phi_2)}$ is used.

Both the 3D model, allowing lateral rotation of dominoes, and the simpler 2D model with flat dominoes constrained to move in $xz$ plane, were studied. In case of small lateral asymmetries in the model, the dominoes exhibited complex 3D motion with lateral rotations (bottom inset in Fig. 6(A)) - curiously, this motion pattern precisely coincides with the one observed in experiment \cite{Destin2018}. Once no lateral asymmetry is introduced, both models give nearly identical results (Fig. 6(B)), therefore, we used a less computationally expensive 2D model -- the dominoes of a single row of spherical particles ($s/l = 0.1$) constrained to move in $xz$ plane.

\begin{figure}
	\begin{center}
		\includegraphics[width=12.5cm]{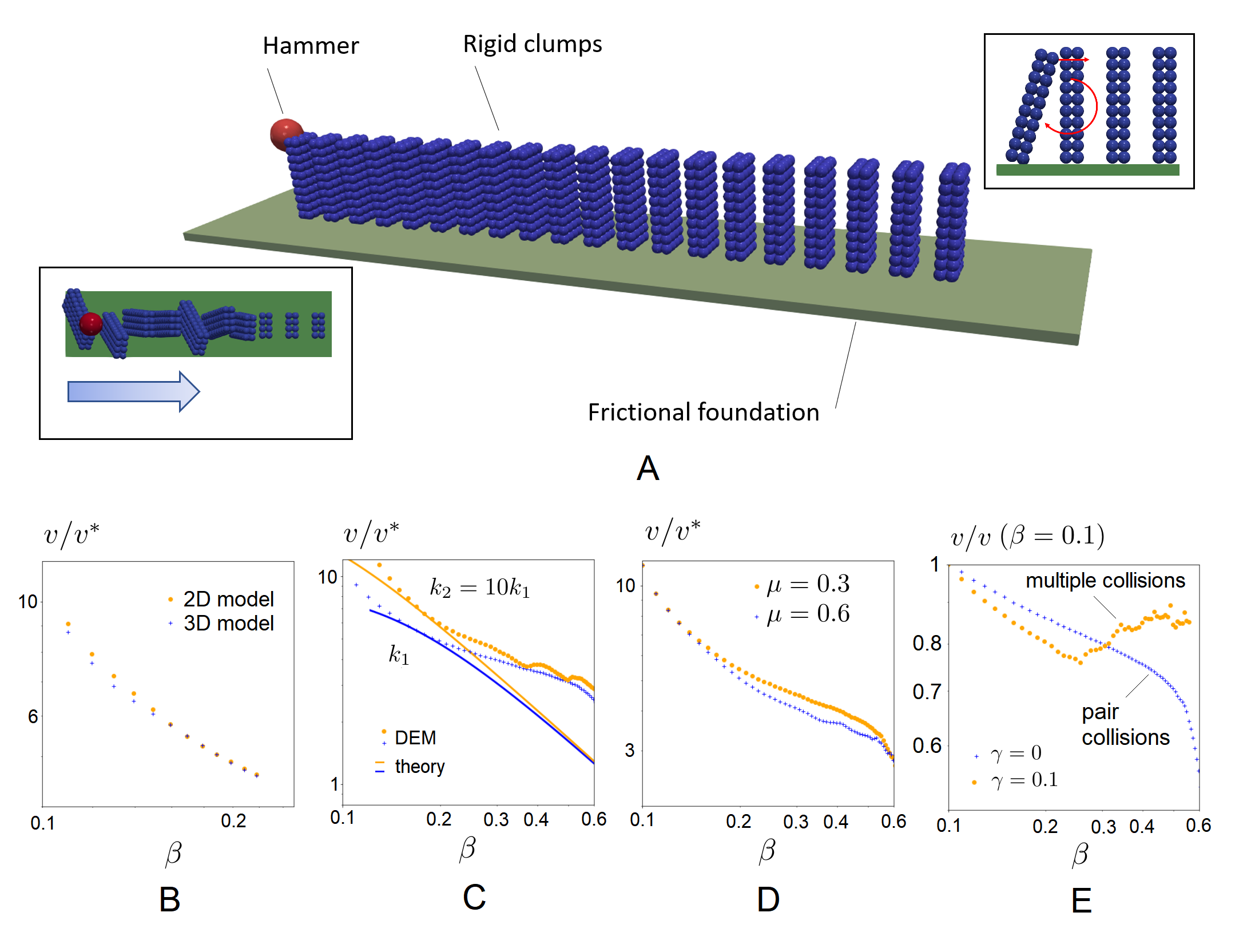}
		\protect\caption{(A) DEM model of the row of dominoes, used in numerical determination of the wave velocity. (B-D) dependence of domino wave velocity on  separation for different model types (B), domino stiffnesses (C) and local damping values (D).}
	\end{center}
\end{figure}

It was established that the predictions of the finite stiffness EJ theory are in qualitative agreement with the DEM simulations (Fig. 6(C)) - stiffness trend is well predicted by the theory. The discrepancies should be associated in the first place with the domino forward rotation during toppling, effectively increasing the velocity. Fig. 6(D) demonstrates that the increased domino-foundation friction ($\mu = 0.6$) reduces this effect. A number of other effects, that are not accounted in a simple analytical model may also affect the domino dynamics, e.g. domino surface roughness in DEM simulations and insufficiently long domino chains used in the model.

Similarly to real experiments \cite{Destin2018} and theoretical predictions (Fig. 5), DEM model exhibited sliding of the domino foundations, featuring forward rotations for small separations (top inset in Fig. 6(A)) and backward rotations/translations for large separations, given sufficiently slender dominoes and small coefficient of friction with the foundation. However, the boundaries presented in Fig. 5(B) were not sharply highlighted by the DEM model. In DEM simulations, the regions in $(F_m/mg, \beta)$ parameter space associated with the pronounced toppling behavior appeared to be significantly larger, though the toppling behavior was almost always accompanied by partial sliding.  This, however, does not contradict the theory, since the latter only predicts the onset of sliding, and not the further domino dynamics.  

Numerical modeling revealed another interesting feature, that was observed in the damped simulations. It appeared that in presence of small amount of acceleration-dependent local damping (please see \cite{Yade_2021} for the exact definition of local damping utilized in our simulation) and in certain range of separation angles the domino $n$ may collide with its neighbor $n-1$ more than once. This effectively increases the toppling propagation velocity (Fig. 6(E)), since every domino gains additional acceleration, not predicted by EJ theory. Similar behavior was observed in case of friction-only damping, but for long enough ($N > 50$) chains. We can see that in presence of energy dissipation the domino $n$, after bouncing back, does not reach its initial vertical position, and starts falling again. This eventually leads to formation of a secondary (collective) wave, which, as can be seen from our simulations, travels faster than the primary pair-collision wave. Once the latter is being overtaken by the former, the wave propagation phenomenon becomes collective, and pair-collision theory is not applicable anymore.   

It is therefore clear that EJ-like behavior can be observed in dissipative systems only for finite-size chains \cite{video_1}, whereas the presence of (very small) energy dissipation leads to onset of the collective propagation mode for a long enough domino chain \cite{video_2}.

\section{Discussion, conclusions and future work}

In our work we presented the mathematical model capable of describing fast elastic waves in domino-like systems, taking into account the finite time of collisions between dominoes. The model avoids the non-physical properties of an instant-collision (EJ) model - infinite contact forces and infinite propagation velocity for the case of small separations. Moreover, an adjusted theory allows to determine the dynamic quantities characterizing collisions, which, in turn, enables establishing the bounds within which the theory is applicable. Our numerical study gives an interesting insight on the role of energy dissipation in the onset of collective (multiple collision) propagation mode.

The adjustment explains the discrepancy with the experimental data noticed previously \cite{Larham_2008} - the regions where the theory diverges from experimental data are simply beyond the borders of the theory's applicability -- either in terms of energy dissipation present in the system, or in terms of admissible, ``no-slip'' domino separations. 

The established bounds limit the applicability of our model to real macroscale domino systems. Collision of real dominoes, especially for large separations, are far from elastic head-on collisions of constrained spherical particles -- the real collision physics involves complex contact geometry, lateral motion, chattering instabilities, non-negligible frictional slip etc (see, e.g. the experimental recordings \cite{Destin2018}). However, our model is very useful in description of some important properties of domino-like discrete mechanical systems. First, it appears that the phenomenon of finite domino wave velocity independent on the initial perturbation does exist in elastic (compliant) domino systems. Second, the existence of pair-collision propagation mode, conditioned by zero or (for finite-sized domino chains) sufficiently low dissipation, makes the total release of the potential energy of toppled dominoes irrelevant for this finite velocity. Third, the finite stiffness of the collisions ensures the continuous elastic wave velocity to be an upper bound for the velocity of the domino-like, discontinuous wave. Finally, the finite-stiffness interactions allow to estimate dynamic/frictional forces in the system and establish the bounds for sequential, pair-wise topping regime of domino wave in terms of relevant system parameters.

The theory can still be refined in few different ways.

1) Numerical modeling indicates, that even in case of slightly inelastic collisions, multiple interactions between sequential dominoes are possible, if the chain is sufficiently long. It would be useful to study collision times for domino-like systems with restitution coefficients close to $1$, and indicate the system parameters leading to multiple interactions within one cycle. 

2) We used quite a simple model of interaction between dominoes - linear contact model, that originates from the earliest works on interactions between discrete elements \cite{Cundall_1979}. This model assumes constant interaction stiffness, whose absolute value is motivated by rather simplistic considerations. Somewhat more detailed analysis may give the refined picture of contact interactions of dominoes.

3) Our analysis does not account for the role of gravity during collision. Gravitational force does non-zero work and apply nonzero torque, that, in principle, should be accounted in the conservation laws.  

4) The analysis uses the simplest possible mass distribution, allowing quick analytical treatment of rotational motion of dominoes. The analysis for more realistic mass distributions will lead to somewhat more complicated expressions for quantities in the conservation laws, and, consequently, for the collision time.   

5) The considerations above are based on the assumption of vanishingly small friction between dominoes, and therefore, absent tangential forces at contact points. As has been discussed by \cite{Stronge_1987}, presence of these tangential forces noticeably affects the scaling of the wave propagation velocity.

6) Fig. 4 demonstrates that in the limit of small separations the domino wave can not exceed the speed of P-wave in the corresponding spring-mass system. However, transition region between the domino wave and P-wave remains unexplored in our work. Clearly, small separations admit emergence and disappearance of elastic contacts between multiple next nearest neighbors in the chain. This effectively creates stiffness nonlinearity, which can result in soliton-like behavior, as discussed in \cite{Nesterenko_1983}. Further exploration of these mechanical behaviors remains outside of the scope of our work.

The generalization 1) can make our theory applicable for slightly non-conservative systems. The generalizations 2) - 5) are expected to result in rather minor adjustments of the quantitative characteristics of the model. The question 6) is an interesting research direction in itself, that can bridge our work with the existing developments within the granular matter community (e.g. \cite{Daraio_2006, Ostanin_2022}).    

Finite collision time domino theory is useful for interpreting experimental results with fast domino-like systems at small scales. We foresee that our results may have relevance for wave propagation in a certain types of discontinuous (granular) soft matter. Moreover, it provides a foundation for modeling these systems with DEM, and helps interpreting the results of such simulations.

The source code of the YADE scripts used in our simulations is available at \url{https://bitbucket.org/iostanin/domino/}. Similar simulation framework was also implemented by the authors in MercuryDPM \cite{weinhart2020fast, Thornton2023}, it is available at \url{https://bitbucket.org/mercurydpm/mercurydpm/}. 

\section*{Acknowledgments}

The work incorporates the the results of few student BS/ARS projects (D.D, C.L, J.W., L.H, L.K, P,B.) accomplished at TFE/ET department of the University of Twente. The assistance from the University of Twente ME BS/ARS program is deeply appreciated. I.O. expresses his gratitude to S. Luding and A. Thornton for the fruitful discussions on the topic.   

\section*{Declarations}

\textbf{Conflict of interest} The authors declare that they have no conflict of interest

\bibliographystyle{unsrtnat}
\bibliography{manuscript}

\begin{thebibliography}{24}
\providecommand{\natexlab}[1]{#1}
\providecommand{\url}[1]{\texttt{#1}}
\expandafter\ifx\csname urlstyle\endcsname\relax
  \providecommand{\doi}[1]{doi: #1}\else
  \providecommand{\doi}{doi: \begingroup \urlstyle{rm}\Url}\fi

\bibitem[Daykin(1971)]{Daykin_1971}
DE~Daykin.
\newblock Falling dominoes, problem 71-19*.
\newblock \emph{SIAM Rev.}, 13:\penalty0 569--570, 1971.

\bibitem[McLachlan et~al.(1983)McLachlan, Beaupre, Cox, and
  Gore]{McLachlan_1983}
BG~McLachlan, G~Beaupre, AB~Cox, and L~Gore.
\newblock Solution for “falling dominoes” (problem 71-19*).
\newblock \emph{SIAM Rev.}, 25:\penalty0 403--404, 1983.

\bibitem[Sun(2020)]{Bohua_2020}
BH~Sun.
\newblock Scaling law for the propagation speed of domino toppling.
\newblock \emph{AIP Advances}, 10\penalty0 (9):\penalty0 095124, 2020.

\bibitem[Song et~al.(2021)Song, Guo, and Sun]{Bohua_2021}
G~Song, X~Guo, and B~Sun.
\newblock Scaling law for velocity of domino toppling motion in curved paths.
\newblock \emph{Open Physics}, 19\penalty0 (1):\penalty0 426--433, 2021.

\bibitem[Bert(1986)]{Bert_1986}
CW~Bert.
\newblock Falling dominoes.
\newblock \emph{SIAM Review}, 28\penalty0 (2):\penalty0 219--224, 1986.

\bibitem[Stronge(1987)]{Stronge_1987}
WJ~Stronge.
\newblock The domino effect: a wave of destabilizing collisions in a periodic
  array.
\newblock \emph{Proc. R. Soc. Lond.}, A409:\penalty0 199–208, 1987.

\bibitem[van Leeuwen(2010)]{vanLeeuwen_2010}
JMJ van Leeuwen.
\newblock The domino effect.
\newblock \emph{American Journal of Physics}, 78\penalty0 (7):\penalty0
  721--727, 2010.

\bibitem[Efthimiou and Johnson(2007)]{Johnson_2007}
C~Efthimiou and MD~Johnson.
\newblock Domino waves.
\newblock \emph{SIAM Rev.}, 49:\penalty0 111--120, 2007.

\bibitem[Shaw(1978)]{Shaw_1978}
DE~Shaw.
\newblock Mechanics of a chain of dominoes.
\newblock \emph{American Journal of Physics}, 46\penalty0 (6):\penalty0
  640--642, 1978.

\bibitem[Shi et~al.(2019)Shi, Liu, Wang, and Liu]{Shi_2019}
T~Shi, Y~Liu, N~Wang, and C~Liu.
\newblock Toppling dynamics of a mass-varying domino system.
\newblock \emph{Nonlinear Dynamics}, 98\penalty0 (3):\penalty0 2261--2275, Nov
  2019.
\newblock ISSN 1573-269X.

\bibitem[Cantor and Wojtacki(2022)]{Cantor_2022}
D~Cantor and K~Wojtacki.
\newblock Effects of friction and spacing on the collaborative behavior of
  domino toppling.
\newblock \emph{Phys. Rev. Appl.}, 17:\penalty0 064021, Jun 2022.

\bibitem[Dalla~Pola et~al.(2023)Dalla~Pola, Darmendrail, Galantay, and
  Müller]{Pola_2023}
L~Dalla~Pola, L~Darmendrail, E~Galantay, and A~Müller.
\newblock Listen! a smartphone inquiry on the domino effect, 2023.

\bibitem[Larham(2008)]{Larham_2008}
R~Larham.
\newblock Validation of a model of the domino effect?
\newblock \emph{arXiv preprint arXiv:0803.2898}, 2008.

\bibitem[Cundall and Strack(1979)]{Cundall_1979}
PA~Cundall and ODL Strack.
\newblock A discrete numerical model for granular assemblies.
\newblock \emph{geotechnique}, 29\penalty0 (1):\penalty0 47--65, 1979.

\bibitem[https://yade dem.org/doc/(2021)]{Yade_2021}
https://yade dem.org/doc/, 2021.

\bibitem[Weinhart et~al.(2020)Weinhart, Orefice, Post, van
  Schrojenstein~Lantman, Denissen, Tunuguntla, Tsang, Cheng, Shaheen, and
  Shi]{weinhart2020fast}
T~Weinhart, L~Orefice, M~Post, MP~van Schrojenstein~Lantman, I~Denissen,
  DR~Tunuguntla, JMF Tsang, H~Cheng, MY~Shaheen, and H~Shi.
\newblock Fast, flexible particle simulations -- an introduction to
  {MercuryDPM}.
\newblock \emph{Computer Physics Communications}, 249:\penalty0 107129, 2020.

\bibitem[Ostanin et~al.(2023)Ostanin, Angelidakis, Plath, Pourandi, Thornton,
  and Weinhart]{Ostanin_2023}
I~Ostanin, V~Angelidakis, T~Plath, S~Pourandi, A~Thornton, and T~Weinhart.
\newblock Rigid clumps in the mercurydpm particle dynamics code.
\newblock 2023.

\bibitem[Podlozhnyuk et~al.(2017)Podlozhnyuk, Pirker, and
  Kloss]{Podlozhnyuk_2017}
A~Podlozhnyuk, S~Pirker, and C~Kloss.
\newblock Efficient implementation of superquadric particles in discrete
  element method within an open-source framework.
\newblock \emph{Computational Particle Mechanics}, 4\penalty0 (1):\penalty0
  101--118, 2017.

\bibitem[Govender et~al.(2016)Govender, Wilke, and Kok]{Govender_2016}
N~Govender, DN~Wilke, and S~Kok.
\newblock Blaze-demgpu: Modular high performance dem framework for the gpu
  architecture.
\newblock \emph{SoftwareX}, 5:\penalty0 62--66, 2016.

\bibitem[https://www.youtube.com/watch?v=9hPIobthvHg(2018)]{Destin2018}
https://www.youtube.com/watch?v=9hPIobthvHg, 2018.

\bibitem[Nesterenko(1983)]{Nesterenko_1983}
VF~Nesterenko.
\newblock Propagation of nonlinear compression pulses in granular media.
\newblock \emph{Journal of Applied Mechanics and Technical Physics},
  24\penalty0 (5):\penalty0 733--743, Sep 1983.
\newblock ISSN 1573-8620.

\bibitem[Daraio et~al.(2006)Daraio, Nesterenko, Herbold, and Jin]{Daraio_2006}
C~Daraio, VF~Nesterenko, EB~Herbold, and S~Jin.
\newblock Tunability of solitary wave properties in one-dimensional strongly
  nonlinear phononic crystals.
\newblock \emph{Phys. Rev. E}, 73:\penalty0 026610, Feb 2006.

\bibitem[Ostanin et~al.(2022)Ostanin, Cheng, and Magnanimo]{Ostanin_2022}
I~Ostanin, H~Cheng, and V~Magnanimo.
\newblock Simulation-guided optimization of granular phononic crystal structure
  using the discrete element method.
\newblock \emph{Extreme Mechanics Letters}, 55:\penalty0 101825, 2022.
\newblock ISSN 2352-4316.

\bibitem[Thornton et~al.(2023)Thornton, Plath, Ostanin, G{\"o}tz, Bisschop,
  Hassan, Roeplal, Wang, Pourandi, and Weinhart]{Thornton2023}
AR~Thornton, T~Plath, I~Ostanin, H~G{\"o}tz, JW~Bisschop, M~Hassan, R~Roeplal,
  X~Wang, S~Pourandi, and T~Weinhart.
\newblock Recent advances in mercurydpm.
\newblock \emph{Mathematics in Computer Science}, 17\penalty0 (2):\penalty0 13,
  Jun 2023.

\end{thebibliography}

\end{document}